\documentclass[11pt]{article}

\columnwidth\textwidth
\usepackage{graphicx}
\usepackage{epsfig}

\begin{document}
\title{The position of the quasielastic peak\\
and\\
electron Coulomb distortion\\
in\\
$(e,e')$ scattering}
\author{Andreas Aste\\
Department of Physics and Astronomy, University of Basel,\\
Klingelbergstrasse 82, 4056 Basel, Switzerland}
\date{November 28, 2006}
\maketitle

\begin{center}
\begin{abstract}
The position of the qua\-si\-elas\-tic peak for $(e,e')$ scattering off
$^{208}$Pb extracted from a selected data set measured at Saclay is related to a
heuristic theoretical description. An analysis of the data shows that the peak position can
be described very accurately by a simple equation in the relevant kinematic
region where a pronounced peak is observable. The simple findings result in
a concluding comment related to recent calculations concerning the
Coulomb distortion in $(e,e')$ scattering for heavy nuclei.
\end{abstract}
\end{center}

\twocolumn
\section{Introduction}

Inclusive $(e,e')$ scattering has several features which make it a very useful
tool for investigating the pro\-per\-ties of nuclei, since the interaction of the
electron with nuclei is well understood in terms of the electromagnetic
and the electroweak interaction.
Inclusive scattering, where only the scattered electron is observed,
provides information on a number of
interesting nuclear properties like, e.g., the nuclear Fermi mo\-men\-tum
\cite{Whitney74}, high-mo\-men\-tum components in nuclear wave
functions \cite{Benhar94a}, modifications of nucleon form factors
in the nuclear medium \cite{Jourdan96a}, the scaling properties
of the qua\-si\-elas\-tic response allow to study the reaction
mechanism \cite{Day90}, and extrapolation of the qua\-si\-elas\-tic
response to infinite nucleon number
$A=\infty$ provides us with a very valuable observable
of infinite nuclear matter \cite{Day89}.
In this paper we focus on the position the qua\-si\-elas\-tic peak,
which is clearly visible when the $(e,e')$ cross section is plotted
versus the energy loss of the scattered electron for an energy range
of the electrons in the region of some few hundred MeV.
For the present analysis, data taken at Saclay \cite{Zghiche94}
for $^{208}$Pb are studied. The presented study also sheds some light
on the problem of Coulomb distortions in $(e,e')$ scattering off
heavy nuclei, which has regained some recent interest.

\section{Position of the qua\-si\-elas\-tic peak}
\begin{figure}[htb]
\begin{center}
        \includegraphics[width=8.0cm]{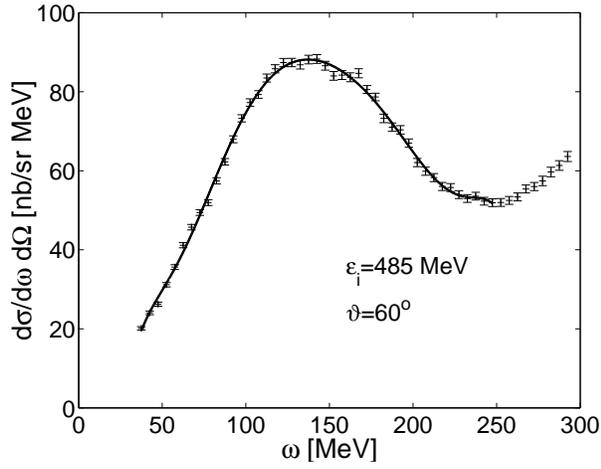}
        \caption{Cross section for $(e,e')$ scattering of a $^{208}$Pb
                 nucleus with initial electron energy $\epsilon_i=485 \, \mbox{MeV}$
                 and scattering angle $\vartheta=60^o$. The plot displays
                 also a least squares tenth order polynomial fit in the
                 qua\-si\-elas\-tic region.}
        \label{saclay}
\end{center}
\end{figure}

\begin{figure}[htb]
\begin{center}
        \includegraphics[width=8.0cm]{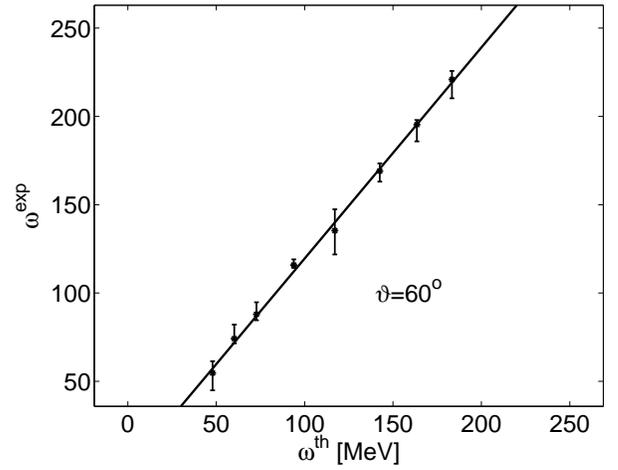}
        \caption{Peak position according to Eq. (\ref{effslope}) versus experimental data
                 for electron scattering angle $\vartheta=60^o$.
                 The data points correspond to initial electron energies of
                 $\epsilon_i=262$, $310$, $354$, $420$, $485$, $550$, $600$,
                 and $645 \, \mbox{MeV}$. The relation is linear, however the slope of the linear
                 fitting function is 1.195.}
        \label{slope60}
\end{center}
\end{figure}

\begin{figure}[htb]
\begin{center}
        \includegraphics[width=8.0cm]{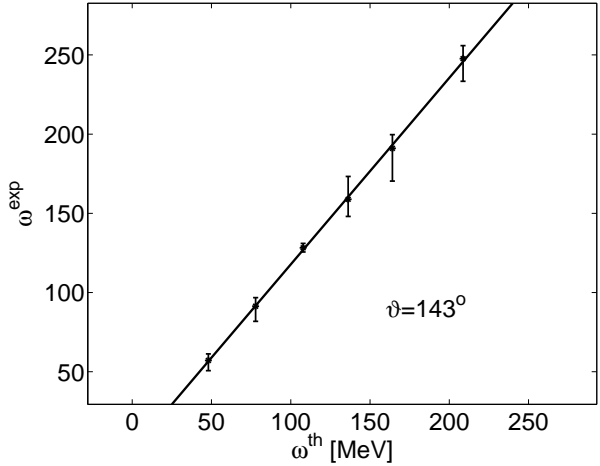}
        \caption{Peak position according to Eq. (\ref{effslope}) versus experimental data
                 for electron scattering angle $\vartheta=143^o$.
                 The data points correspond to initial electron energies of
                 $\epsilon_i=140$, $206$, $262$, $310$, $354$, and
                 $420 \, \mbox{MeV}$. The slope of the linear
                 fitting function is 1.178.}
        \label{slope143}
\end{center}
\end{figure}

From a simplified point of view, qua\-si\-elas\-tic $(e,e')$ scattering may be described
as the scattering process of an electron off all the individual nucleons constituting
the nucleus. Many theoretical calculations for inclusive scattering in connection
with the problem of Coulomb distortions, which are presently available, rely on this
simplified picture, and are correspondingly based on single particle shell model descriptions.
The width of the peak is mainly due to the Fermi motion of the nucleons inside the nucleus,
whereas the position of the peak can be approximately inferred from a simple
classical consideration. One first assumes that the electron with
initial and final momentum $\vec{k}_i$ and $\vec{k}_f$ knocks a
nucleon at rest inside the fixed nucleon, transferring thereby an energy $\omega$
to the nucleon, which is given by the difference of the initial and final electron
energy $\epsilon_i-\epsilon_f$, with $\epsilon_{i,f}=|\vec{k}_{i,f}|$ for highly
relativistic electrons in the present case. The heavy nucleus acts solely as a spectator
within this simplified picture.
The three-momentum transferred to the nucleon is $\vec{q}=\vec{k}_i-\vec{k}_f$,
such that from energy conservation follows
\begin{equation}
\omega=(\vec{p}_f^{\, 2}+m_n^2)^{1/2}-m_n=(\vec{q}^{\, 2}+m_n^2)^{1/2}-m_n, \label{eq1}
\end{equation}
where $m_n$ is the (in-medium) mass of the nucleon and $\vec{p}_f$ the momentum of the
knocked nucleon. Squaring Eq. (\ref{eq1}) one obtains
\begin{equation}
Q^2=\vec{q}^{\, 2}-\omega^2=2 m_n \omega, \quad \mbox{or} \quad \omega=\frac{Q^2}{2 m_n}.
\end{equation}
The four-momentum transfer squared $Q^2$ itself is $\omega$-dependent. Therefore, if the electron
scattering angle $\vartheta$ is given, one may write in a more explicit form
\begin{equation}
\omega = \frac{\epsilon_i^2 (1 - \cos \vartheta)}{m_n + \epsilon_i (1-\cos \vartheta)}.
\end{equation}
However, the energy of the nucleons that leave the nucleus is not given by the energy
loss of the electron, but is reduced by a removal energy of the nucleon.
An more ambitious expression for the position of the qua\-si\-elas\-tic
peak, which can be often found in the literature, is
\begin{equation}
\omega=\frac{Q^2}{2 m_n} + \bar{E}_{rem},
\end{equation}
where $\bar{E}_{rem}$ is an average nucleon removal energy.
A slightly modified expression for the position of the
qua\-si\-elas\-tic peak follows
\begin{equation}
\omega = \frac{\epsilon_i^2 (1 - \cos \vartheta)+m_n \bar{E}_{rem}}{m_n +
\epsilon_i (1-\cos \vartheta)} \label{noeff}.
\end{equation}
Finally, one may try to improve expression Eq. (\ref{noeff}) further by replacing
the initial electron energy by an effective value $\epsilon'_i=\epsilon_i
+20 \, \mbox{MeV}$, since the average enhancement of the kinetic energy of the electron
inside the nuclear region due to the strong attractive Coulomb potential is
given by $\sim 20 \, \mbox{MeV}$ in the case of lead, such that we arrive at a
simple heuristic equation for the theoretical position of the qua\-si\-elas\-tic peak
\begin{equation}
\omega^{th} = \frac{{\epsilon'}_i^2 (1 - \cos \vartheta)+m_n \bar{E}_{rem}}{m_n +
\epsilon'_i (1-\cos \vartheta)}. \label{eff}
\end{equation}
A typical example for a qua\-si\-elas\-tic peak is shown in Fig. \ref{saclay}, where
Saclay data for $^{208}$Pb \cite{Zghiche94} are displayed for initial
electron energy $\epsilon_i=485 \, \mbox{MeV}$ and scattering angle
$\vartheta=60^o$.
For this scattering angle, a qua\-si\-elas\-tic peak is visible 
for initial electron energy
$\epsilon_i=262$, $310$, $354$, $420$, $485$, $550$, $600$, and $645 \, \mbox{MeV}$,
whereas for a scattering angle of $143^o$, the
experimental data display a peak for $\epsilon_i=140$, $206$, $262$, $310$, $354$, and
$420 \, \mbox{MeV}$. The position of each peak and the corresponding
inaccuracy of the peak position was extracted from the data by the help of
polynomial least squares fits with varying order.
However, if one plots the theoretical position Eq. (\ref{eff}) of the corresponding peaks against
the measured positions with a typical nucleon mass of $m_n=939 \, \mbox{MeV}$,
one finds by a least squares fit that the position of the
peaks is described much more accurately by
\begin{equation}
\omega^{th} = \sigma \frac{{\epsilon'}_i^2 (1 - \cos \vartheta)+m_n \bar{E}}{m_n +
\epsilon'_i (1-\cos \vartheta)} \label{effslope}
\end{equation}
with $\sigma=1.195$ and $\bar{E}=12.8 \, \mbox{MeV}$
in the case of $\vartheta=60^o$, and $\sigma=1.178$, $\bar{E}=13.7 \, \mbox{MeV}$
for $\vartheta=143^o$ (see Figs. \ref{slope60} and \ref{slope143}).
In both cases, $\bar{E}$ is even smaller than the average binding
energy of a nucleon in a lead nucleus \cite{Batenburg} and $\bar{E}$ should be considered
rather as a phenomenological fitting parameter and not necessarily as a removal energy.

\begin{figure}[htb]
\begin{center}
        \includegraphics[width=8.0cm]{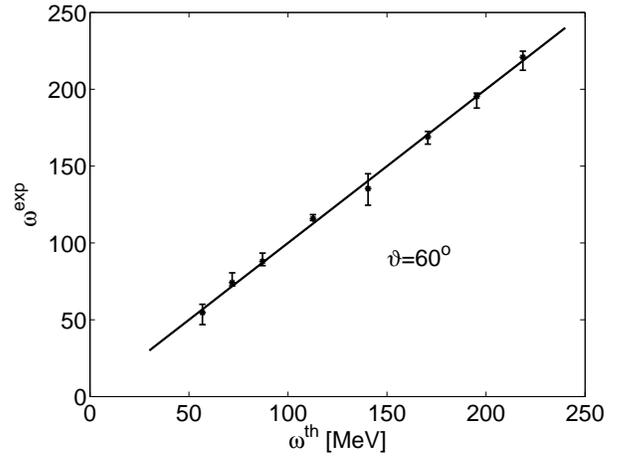}
        \caption{Peak position according to Eq. (\ref{effslope2}) versus experimental data
                 for electron scattering angle $\vartheta=60^o$.}
        \label{effective60}
\end{center}
\end{figure}

\begin{figure}[htb]
\begin{center}
        \includegraphics[width=8.0cm]{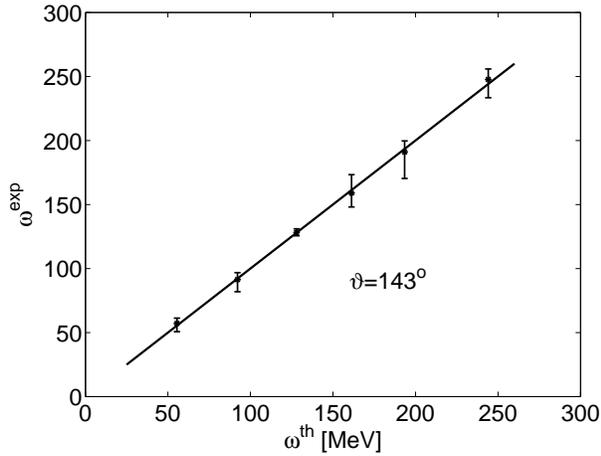}
        \caption{Peak position according to Eq. (\ref{effslope}) versus experimental data
                 for electron scattering angle $\vartheta=143^o$.}
        \label{effective143}
\end{center}
\end{figure}

The observation above automatically leads to the idea that one may,
instead of introducing a factor $\sigma$ in Eq. (\ref{effslope}),
change the nucleon mass into an effective mass parameter $\tilde{m}_n$.
A fit of the experimentally measured peak positions to  
\begin{equation}
\omega^{th} = \frac{{\epsilon'}_i^2 (1 - \cos \vartheta)+\tilde{m}_n \bar{E}}{\tilde{m}_n +
\epsilon'_i (1-\cos \vartheta)} \label{effslope2}
\end{equation}
with $\tilde{m}_n$ and $\bar{E}$ as fitting parameters is shown in Figs.
\ref{effective60} and \ref{effective143}. The experimental data are reproduced
in a very satisfactory manner by Eq. (\ref{effslope2})
for $\vartheta=60^o$ with $\tilde{m}_n=721 \, \mbox{MeV}$
and $\bar{E}=12.8 \, \mbox{MeV}$ and for $\vartheta=143^o$ with
$\tilde{m}_n=662 \, \mbox{MeV}$ and $\bar{E}=10.0 \, \mbox{MeV}$.

There is no need to interpret the effective mass parameter $\tilde{m}_n$ within a physical
picture. It simply incorporates in an efficient way the complex interaction processes
that are taking place inside the nucleus, i.e. the interaction of the
knocked effective nucleon with the other nucleons inside the nucleus or even
effects like pion production which becomes important at higher energy transfer.
In this sense, the mass parameter $\tilde{m}_n$ should not be related directly to the effective
(momentum transfer dependent) nucleon mass as it is investigated in \cite{Zghiche94},
although it is clear that the size of $\tilde{m}_n \sim 0.7 \ldots 0.8 \, m_n$ is quite typical.
Naturally, the above description breaks down at high energy transfer, when the quasielastic
peak is no longer visible in the $(e,e')$ cross section data. One may mention \cite{Hotta} as
another early work which addressed the position of the quasielastic peak in a more theoretical
framework.

\section{Coulomb corrections}
This section comments on calculations of the Ohio group \cite{Jin1}
and the Basel group \cite{Basel1}, which dealt with the problem
of Coulomb distortion in the case of the highly charged $^{208}$Pb
nucleus.

We first comment on Fig. 4 in \cite{Jin1}. The left part of the figure
displays theoretical calculations for $(e,e')$ scattering cross sections
with $\epsilon_i=310 \, \mbox{MeV}$ and $\vartheta=143^o$. The peak of the
plane wave Born approximation (PWBA) curve is located at $\omega=140 \, \mbox{MeV}$,
whereas the peak of the effective momentum approximation (EMA) is located at
$\omega=163 \, \mbox{MeV}$. However, the EMA as it is applied in \cite{Jin1} is a plane wave
calculation with the initial electron energy of $310 \, \mbox{MeV}$ replaced
by an effective electron energy of $310\, \mbox{MeV} - V(0)=
335 \, \mbox{MeV}$, combined with a renormalization
of the cross section by a constant focusing factor depending on the initial
electron energy only, such that the position of the peak is not influenced by this factor.
Furthermore, the energy transfer $\omega$ is left unchanged in the formal calculations,
such that, e.g.,
$Q^2 = \vec{q}^{\, 2}-\omega^2 =2 \epsilon_i \epsilon_f (1-\cos \vartheta)$
is replaced by
$Q^2_{eff}=2 (\epsilon_i+25 \, \mbox{MeV})(\epsilon_i -\omega + 25 \, \mbox{MeV})
(1-\cos \vartheta)$.
The focusing factor for the final state electron wave function is not neglected
in this approach, but already absorbed in the enhanced phase space factor of the final
state electron. $V(0)=- 25 \, \mbox{MeV}$ is the electrostatic potential energy of the
electron in the center of the nucleus. There exist different descriptions of how to apply the
EMA, which are all, of course, equivalent (see, e.g., also the introduction in \cite{Gueye}
for an alternative description). We will argue below that the EMA is in fact a valuable
tool to calculate Coulomb effects in $(e,e')$ scattering, however, an average
potential value $\bar{V} \simeq -19 \, \mbox{MeV}$ should be used instead of
the \emph{ad hoc} value $V(0)$.

The essence of the EMA is to account for two effects of the attractive Coulomb field of the
nucleus. First, the electron wave function is focused towards the nuclear region, and second, the
electron momentum is enhanced due to the attractive Coulomb force. Both the focusing and
the momentum of the electron vary inside the nucleus, but the average effect can be
described by the average (or effective) Coulomb potential of the nucleus which is given by
$\bar{V} \simeq -19 \, \mbox{MeV}$ for $^{208}$Pb. It has already been observed by Rosenfelder
\cite{Rosenfelder},
that for high-energy electrons, the distorted electron wave can
be approximated by ($\psi_0$ is the constant spin-dependent Dirac spinor)
\begin{equation}
\psi_{\vec{k}}(\vec{r})=\frac{|\vec{k}_{eff}|}{|\vec{k}|} \psi_0 e^{i \vec{k}_{eff} \vec{r}},
\label{rosen}
\end{equation}
with $|\vec{k}_{eff}|=|\vec{k}|-\bar{V}$ for both the initial and final electron momentum.
It should be noted here that Rosenfelder explicitly mentioned already in \cite{Rosenfelder}
that $\bar{V}$ is close to the \emph{mean value} of the electrostatic potential of the nucleus.
Unfortunately, he then wrote down the explicit expression for the \emph{central}
value
\begin{equation}
V(0)=-\frac{3 Z \alpha}{2R}
\end{equation}
of the potential of a homogeneously charged sphere with radius $R$, which is related to the
corresponding mean value by $\bar{V}=4 V(0)/5$.
This minor, but not irrelevant misapprehension for the case of the EMA,
has propagated in the literature since then (see, e.g., \cite{Hotta}).

With a little abuse of notation, we will disregard the negative
sign of $V(0)$ and $\bar{V}$ in the following.
According to the considerations above in connection with Eq. (\ref{effslope2}),
the PWBA peak should be located approximately at ($\tilde{m}_n=662 \, \mbox{MeV}$,
$\bar{E}=10 \, \mbox{MeV}$)
\begin{displaymath}
\omega_{PWBA} \simeq
\frac{{310 \, \mbox{MeV}}^2 (1 - \cos 143^o)+\tilde{m}_n \bar{E}}{\tilde{m}_n +
310 \, \mbox{MeV} (1-\cos 143^o)}
\end{displaymath}
\begin{equation}
\simeq 147.2 \, \mbox{MeV},
\end{equation}
and the EMA peak at
\begin{displaymath}
\omega_{EMA} \simeq
\frac{{335 \, \mbox{MeV}}^2 (1 - \cos 143^o)+\tilde{m}_n \bar{E}}{\tilde{m}_n +
335 \, \mbox{MeV} (1-\cos 143^o)}
\end{displaymath}
\begin{equation}
\simeq 164.9 \, \mbox{MeV}.
\end{equation}
The theoretical (model dependent) and experimental peak
positions do not have to coincide extremely well, still one finds that they agree
in a satisfactory way. On the other hand, the \emph{distance} $\omega_{EMA}-\omega_{PWBA}$
between the peaks is a robust quantity, since it depends only on a comparably small
change of the momentum transfer $Q^2$ into $Q_{eff}^2$, and the behavior of the
peak position as a function of the momentum transfer shows a universal behavior and
is obviously well under control.
Therefore, one should expect a peak shift of
$\Delta \omega = 164.9-147.2 = 17.7 \, \mbox{MeV}$.
However, the peak shift of $23 \, \mbox{MeV}$ displayed in \cite{Jin1}
would rather correspond to an effective potential
of $32.5 \, \mbox{Mev}$. If a free nucleon mass of $939 \, \mbox{MeV}$
was used in the theoretical calculations in conjunction with adapted removal energies
or nuclear potentials in order to obtain the correct position of the peaks, the situation is
worse, since the expected peak shift $\simeq Q^2_{eff}/2 m_n -Q^2/2m_n$
would then correspond to $V(0) \simeq 37 \, \mbox{MeV}$.
Interestingly, the distance between the peaks of the Coulomb corrected DWBA curve,
where exact electron wave
functions have been used for the calculation, and the PWBA curve, is given by $16.5 \, \mbox{MeV}$,
corresponding to an effective energy shift of $\sim 23 \mbox{MeV}$, or, if vacuum nucleon masses
were used, to $V(0) \sim 18 \, \mbox{MeV}$. The definition of the nucleon current in
\cite{Jin1} indeed suggests that the free nucleon mass has been used. 
The figure displays an EMA curve with a peak shift against the PWBA curve
which is obviously too large, but the shift of the DWBA curve is compatible with the EMA.
But there remains a problem with the normalization of the theoretical data.

The situation is harder to interpret for the left plot with theoretical results for
$\epsilon_i=485 \, \mbox{MeV}$ and $\vartheta=60^o$.
Magnifying the plot one finds that the PWBA and the DWBA peaks
are separated only by approximately $5 \, \mbox{MeV}$, compatible with an EMA calculation
using an electrostatic potential value of $V(0) \simeq 15 \, \mbox{MeV}$,
if a nucleon mass of of $939 \, \mbox{MeV}$ is used (otherwise, the situation is worse).
The separation of the PWBA and the EMA peak is again too large and corresponds to an EMA
calculation with $V(0) \simeq 30 \, \mbox{MeV}$ (for $\tilde{m}_n=721 \,
\mbox{MeV}$) or even $V(0) \simeq 39 \, \mbox{MeV}$ (for $m_n=939 \, \mbox{MeV}$).

Again, the figure supports rather an EMA type behavior of the DWBA results contrary
to the initial intention of \cite{Jin1}, but there is again a problem with the normalization.
The observations above have lead to the attempt to describe the theoretical DWBA results
presented in \cite{Jin1} by an improved EMA, where the normalization of EMA results was
modified in a suitable manner \cite{Basel2}.

We finally comment on the eikonal calculations (EDWBA) presented in \cite{Basel1},
which originally seemed to be compatible with the results in \cite{Jin1}.
In this paper, Coulomb corrected cross sections were calculated based
on the eikonal approximation, where the distortion of the electron wave functions
is described by the help of an eikonal integral, which is easily accessible
by numerical calculations. It should be pointed out that
the paper contains a pedagogical introduction, which may
mislead to the assumption that the electron current and matrix elements
(like Eq. (40) in \cite{Basel1}) were calculated within a simplified Klein-Gordon model, however,
the full spinor formalism was applied to electrons within the eikonal framework \cite{Tjon}.
The PWBA and EDWBA curves in Fig. 3 of \cite{Basel1} show a indeed similar behavior as the 
PWBA and DWBA results in \cite{Jin1}.
\begin{figure}[htb]
\begin{center}
        \includegraphics[width=8.0cm]{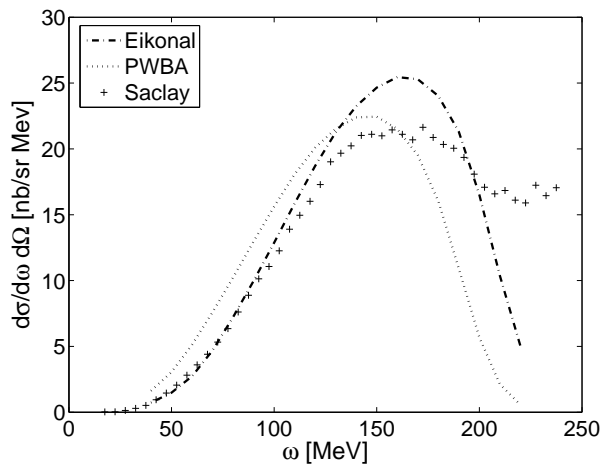}
        \caption{Cross sections for $(e,e')$ scattering with $\epsilon_i=310 \, \mbox{MeV}$
                 and $\vartheta=143^o$ obtained in \cite{Basel1}. The plots displays a similar behavior
                 as the results presented in \cite{Jin1}, however, the Coulomb corrected eikonal cross
                 sections are too large due an overestimation of the electron wave function focusing.}
        \label{eiko143}

\end{center}
\end{figure}
As in \cite{Jin1}, the Coulomb corrected cross section is higher at the peak than
in the PWBA case, seemingly in contradiction with the EMA.
However, there is a simple explanation for this discrepancy.
For the EDWBA calculations in \cite{Basel1}, a focusing value for the electron
cross section was used which corresponds basically to the central value of the electrostatic
potential, i.e. $V(0)=25 \, \mbox{MeV}$. But exact solutions of the Dirac equation
for electrons in the electrostatic potential of a $^{208}$Pb nucleus reveal that the
average focusing inside the nucleus is lower than the central value, and can be
calculated reliably from an effective potential value of approximately $19-20 \, \mbox{MeV}$
\cite{Basel3}.
This lead to an overestimation of the Coulomb corrected cross sections.
Note that the average momentum of the electrons is well described by the eikonal
integral and corresponds also to an average value of approximately $19 \, \mbox{MeV}$.

\begin{figure}[htb]
\begin{center}
        \includegraphics[width=8.0cm]{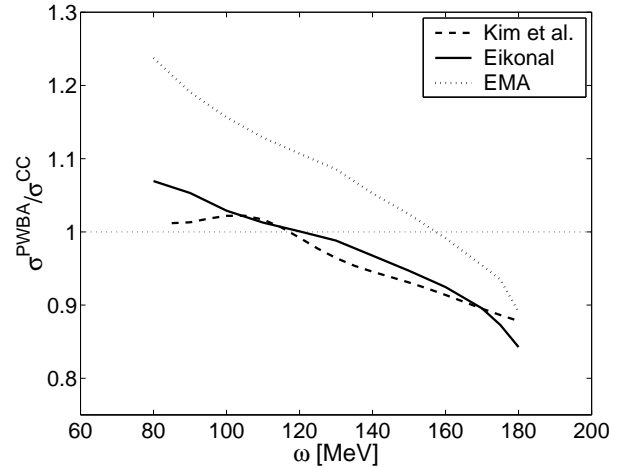}
        \caption{Comparison of Coulomb corrections ($\epsilon_i=485 \, \mbox{MeV}$,
                 $\vartheta=60^o$) for different approaches in \cite{Basel1}. For the EMA,
                 an effective potential value of $25 \, \mbox{MeV}$ was used.}
        \label{ratio_60}
\end{center}
\end{figure}

\begin{figure}[htb]
\begin{center}
        \includegraphics[width=8.0cm]{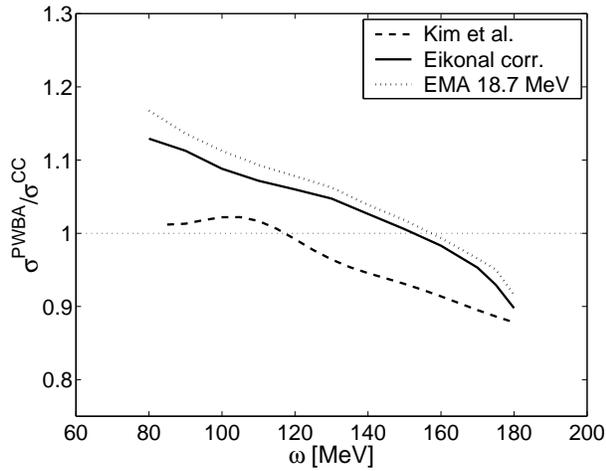}
        \caption{The same as in Fig. \ref{ratio_60}, but with corrected focusing
                 in the eikonal calculation and an EMA curve obtained by using an average
                 potential $\bar{V}=18.7 \, \mbox{MeV}$.}
        \label{ratio_60corr}
\end{center}
\end{figure}

Therefore, the results presented in \cite{Basel1} have to be corrected in two steps,
with the result illustrated below for Fig. 4 in \cite{Basel1}, which is displayed
again as Fig. \ref{ratio_60} in this paper. The figure shows the ratio of the cross sections,
calculated in PWBA, with the Coulomb corrected cross sections according to the EDWBA and EMA.
First, the EMA curve has to be recalculated
for an effective potential of $19 \, \mbox{MeV}$ instead of $25 \, \mbox{MeV}$.
This slightly reduces the ratio $\sigma_{PWBA}/\sigma_{EMA}$ and moves the corresponding
dotted curve closer to one (the horizontal line). Second, the focusing factors
of the EDWBA calculations must be corrected. As an example, for $\epsilon_i=485 \, \mbox{MeV}$
and $\epsilon_f=385 \, \mbox{MeV}$, the original focusing factor was given by
\begin{equation}
f=\frac{(485+25)^2 \times (385+25)^2}{485^2 \times 385^2} \simeq 1.254. \label{example1}
\end{equation}
The correct focusing factor should rather be
\begin{equation}
f_{corr}=\frac{(485+19)^2 \times (385+19)^2}{485^2 \times 385^2} \simeq 1.189. \label{example2}
\end{equation}
Accordingly, the EDWBA cross section has to be reduced by
$5.2 \%$ at $\omega=100 \, \mbox{MeV}$
and the corresponding solid curve moves upwards in the plot.
The result is shown in Fig. \ref{ratio_60corr}. Note that for Fig. \ref{ratio_60corr},
also locally varying focusing factors obtained from exact solutions of the Dirac
equation were used in conjunction with the eikonal approximation for the phase
of the electron wave functions. An attempt to calculate corrections to the focusing
near the nuclear center has already been presented in \cite{Knoll}, which, however,
does not lead to reliable predictions in the important surface region of the nucleus.
The results for a calculation using a locally varying focusing
or a calculation using a corresponding, but constant mean focusing, do not differ
significantly. Therefore, the calculational examples Eqns. (\ref{example1}) and
(\ref{example2}) contain the main essence of the present consideration, i.e.,
that the focusing used in \cite{Basel1} corresponds to an effective potential of
$25 \, \mbox{MeV}$ which is too large. For the EMA calculation presented in Fig.
\ref{ratio_60corr}, a slightly smaller value $\bar{V}=18.7 \, \mbox{MeV}$ than in the
calculational example above was used, since this value has been determined experimentally
to be $\bar{V}=18.7 \pm 1.5 \, \mbox{MeV}$ for $^{208}$Pb \cite{Gueye}.

One observes that the EMA and the EDWBA agree very well,
if the correct effective potential and focusing factors are used.
This result was also obtained in \cite{Basel3}, where exact solutions of the Dirac
equation were used in conjunction with a simplified model for the nuclear current.
Detailed calculations with realistic nuclear current will be presented in a forthcoming
paper in the near future. Note that a small discrepancy of the order of $1-2 \%$
as in Fig. \ref{ratio_60corr} between the EMA and the EDWBA still leaves the possibility
for an improved EMA as proposed in \cite{Basel2}, however, such an attempt to
make the EMA perfect must be based on full DWBA calculations. Detailed DWBA calculations
will also be necessary in order to find optimal EMA values for $\bar{V}$ and the related
effective kinematic variables for different kinematic settings.

One has to conclude that using exact Dirac wave functions or an eikonal approximation
with correct normalization of the wave function in the nuclear region confirm the
usefulness of the EMA, if an appropriate effective potential is
used in the kinematic regions considered in this work.
Additionally, the energy of the final state electron should be larger than
$200 \, \mbox{MeV}$ and the momentum transfer $Q^2$ larger than $(300 \, \mbox{MeV})^2$ \cite{Basel3},
such that the length scale of the electron wave functions and of the exchanged virtual photon is
sufficiently small compared to the size of the nucleus and a semiclassical
behavior sets in. It is a pleasant fact that the focusing factors of the electron wave functions
and the modification of the momentum transfer due to the attractive
Coulomb potential can be calculated from the same effective potential
$\bar{V}$. It must be pointed out that this
fact is to some extent accidental, since it is only typical for charge distributions which are
close to the charge distribution of a homogeneously charged sphere.
In a more ambitious approach to the EMA, one would use two different $k_{eff}$'s in
Eq. (\ref{rosen}).

\end{document}